\documentclass[a4paper]{article}
\usepackage[comma,numbers,square,sort&compress]{natbib}
\usepackage{amsfonts,amssymb,bm,amsmath}
\usepackage{epstopdf}
\usepackage{graphicx} 
\usepackage{float}

\usepackage{pst-node}
\usepackage{subfigure,pgf,tikz,pst-node}
\usetikzlibrary{arrows,shapes,automata,positioning,fit}
\usepackage{epsfig}
\usepackage[justification=centering]{caption}

\newcommand{\lm}{\lambda}

\newcommand{\R}{\mathbb{R}}
\newcommand{\N}{\mathcal{N}}
\newcommand{\G}{\mathcal{G}}
\newcommand{\I}{\mathbb{I}}

\newtheorem{assumption}{Assumption}
\newtheorem{lemma}{Lemma}
\newtheorem{theorem}{Theorem}
\newtheorem{remark}{Remark}
\newtheorem{problem}{Problem}
\newtheorem{corollary}{Corollary}

\newcommand{\pb}{\noindent\textbf{Proof. } }
\newcommand{\pe}{\hfill\rule{4pt}{8pt}}

\def\rm{\mathrm}
\begin{document}

\title{Robust Positive Consensus for Heterogeneous Multi-agent Systems}

\author{Ruonan Li, Yutao Tang, and Shurong Li \footnote{This work was supported by National Natural Science Foundation of China under Grants 61973043. Ruonan Li, Yutao Tang, and Shurong Li are all with the School of Artificial Intelligence,  Beijing University of Posts and Telecommunications, Beijing, China (e-mails: nanruoliy@163.com, yttang@bupt.edu.cn, lishurong@bupt.edu.cn).}}

\date{}

\maketitle

{\noindent\bf Abstract}: This paper investigates a robust positive consensus problem for a class of heterogeneous high-order multi-agent systems subject to external inputs. Compared with existing multi-agent consensus results, the most distinct feature of the formulated problem is that the state variables of all heterogeneous agents are confined in the positive orthant. To solve this problem, we present a two-step design procedure. By constructing an auxiliary multi-agent system as positive local reference generators, we incorporate the reference generator into some applicable decentralized robust tracking controller for each agent. The proposed distributed algorithm is proven to ensure a robust consensus fulfilling certain prescribed pattern for the multi-agent system under switching topology in the sense of finite-gain stability with respect to the external inputs. A simulation example is finally given to illustrate the effectiveness of our design. 

\bigskip


\section{Introduction}

Over the past decades, there has been a tremendous expansion of the research on the multi-agent coordination problem due to its wide applications in sensor networks, robotics, and power systems. Particularly, the fundamental consensus problem has been intensively studied and generalized for various kinds of agent dynamics from integrators to general linear systems and also typical classes of nonlinear ones under different communication topologies, to name a few, \cite{ren2008consensus, hong2006tracking, ni2010leader,su2011cooperative,tang2016ijss, wang2016robust, liu2018distributed, zhang2022optimal,an2022optimal,cao2022ETCdelay, Kaviarasan2021tracking}.

In practical applications arising in the areas of chemical process industry, electronic circuit design, communication networks,  and biology, we may face an important class of multi-agent systems composed of a group of positive subsystems. { Different from standard multi-agent systems, the state variables of positive agents are confined to the positive orthant. Although we might view the whole positive multi-agent systems as a single but large-scale positive systems, the coordination of positive multi-agent systems suffers from at least two extra difficulties compared with conventional designs for a single positive plant \cite{farina2000positive, liu2010timedelay, blanchini2015switched, Li2021practical, yang2021proportional}.} First, the controller for each agent (as an individual input channel) is not allowed to use the full state of the whole multi-input multi-output systems and the information flow among the agents should be compatible with some prior (time-varying) structure.  Second, we have to ensure the positivity of the state variables of each agent, which might have different dimensions and be affected by others.  These two aspects together make the coordination problem of multiple positive systems much more challenging. { As a consequence, compared with the intensive research for conventional standard multi-agent systems,  distributed coordination results for general positive multi-agent systems are relatively few.}  

Meanwhile, integrators are typical positive systems. Thus, the classical integrator-type multi-agent systems are naturally interconnected positive multi-agent systems. Although the positive constraint is not required in designing rules for single-integrator multi-agent systems, the fundamental consensus dynamics are indeed positive. In \cite{valcher2013stabilizability}, such a positive constraint was explicitly discussed in the consensus problem for some positive linear multi-agent systems.  Necessary and sufficient conditions for the consensusability of all agents were derived under the positive constraint.  Note that the positive consensus in this work was achieved via solving a positive static output feedback stabilization problem for the whole multi-agent system. Further efforts along this technical line have been made in \cite{valcher2017consensus, sun2017stabilization, liu2019positivity} using state feedback. Moreover, some authors considered dynamic rules to relax the required conditions, e.g., \cite{wu2018observer, yang2019positive, bhattacharyya2022positive}.   However, all these positive consensus results are only derived for homogeneous multi-agent systems where all agents share the identical dynamics. Noticing  the various kinds of conventional consensus results for different classes of multi-agent systems, it is natural for us to ask whether and how the positive consensus problem can be solved for heterogeneous multi-agent systems.

On the other hand, the agents' dynamics may not be perfectly known due to various uncertainties from either inaccurate modeling or environmental disturbances. Note that the controllers designed for nominal dynamics might fail to ensure the same performance for physical plants and probably cause some instability issues. Thus, it is crucial to take the robustness of controllers into account when solving the multi-agent coordination problems. Although many interesting results have been delivered for single positive systems in the literature, e.g., \cite{shen2016static,li2015stability,xu2022positive}, it is not clear how to extend these results to positive multi-agent systems for a robust consensus.

Based on the aforementioned observations, we will focus on a group of heterogeneous positive linear multi-agent systems  where the agent dynamics are allowed to be different from each other in both system matrices and the dimension of state spaces. Moreover, we assume the agents are subject to external inputs in the dynamics. We aim at distributed rules for these positive agents to reach an output consensus corresponding to some predefined pattern with their state variables being positive. Since the agents' dynamics are high-order and subject to external inputs, we expect that an exact output consensus can be asymptotically ensured for the nominal multi-agent systems while such performance is robust with respect to these external inputs. Note that we have to achieve the expected robust consensus goal and ensure the positive constraints for each agent simultaneously. Hence the considered robust positive consensus problem for these heterogeneous positive agents has some exclusive challenges in contrast to existing (robust) consensus results for standard linear multi-agent systems or positive consensus for homogeneous multi-agent systems.

To overcome the difficulties brought by the positive constraint and heterogeneous uncertain agent dynamics, we constructively present a two-step procedure. First, we construct an auxiliary multi-agent system as a positive local reference generator for each agent. Then, we design an effective robust tracking controller and bring the two parts together with rigor solvability analysis to solve the robust positive consensus problem. The contributions of this paper can be summarized as follows:
\begin{itemize}
	
	\item A robust positive consensus problem is formulated for a group of high-order positive multi-agent systems subject to external inputs. Our problem is an extended version of existing positive consensus results with exactly known agent dynamics \cite{liu2019positivity,valcher2017consensus,bhattacharyya2022positive}.  
	\item  Compared with existing positive consensus results for homogeneous multi-agent systems,  both state and output feedback controllers are provided to remove the identical agent dynamics requirement and allow the consensus trajectory to meet some prespecified pattern including the finite constant as special cases. 
\end{itemize}

The rest of this paper is organized as follows. We first introduce some preliminaries on our notations and positive system in Section \ref{sec:pre}. Then we present the formulated robust positive consensus problem in Section \ref{sec:form}. The main results are given in Section \ref{sec:main}. 
We also provide a simulation example to illustrate the effectiveness of our algorithms in Section \ref{sec:simulation} along with conclusions in Section \ref{sec:conclusion}.

\section{Preliminary}\label{sec:pre}

In this section, we introduce some preliminaries on our notations and positive systems.

Let $\R^{n}$ be the $n$-dimensional real space. Denote by $\R^{n \times  m}$ the set of all $n \times m $ matrices with entries in $\R$. Let $\I_n$
be the $n$-dimensional identity matrix. 
${\bf 1}$ (or ${\bf 0}$) denotes an all-one (or all-zero) matrix or vector with
proper dimensions. $\mbox{col}(a_1,\,{\dots},\,a_n) = {[a_1^\top,\,{\dots},\,a_n^\top]}^\top$ for column vectors  $a_i\; (i=1,\,{\dots},\,n)$ with compatible dimensions. Let $\mbox{diag}(b_1,\,{\dots},\,b_n)$ represent an $n\times n$ diagonal matrix with diagonal elements $b_1,\,\dots,\, b_n$. For matrices $B_1,\,\dots,\,B_n$,    $\mbox{blkdiag}(B_1,\,{\dots},\,B_n)$ represents the block diagonal matrix with diagonal blocks $B_1,\,\dots,\,B_n$. For a vector $x$ (or matrix $A$) , $\|x\|$ (or $\|A\|$) denotes its Euclidean (or spectral) norm.   

Let $\R_+^n$ be the nonnegative orthant.  Denote the set of all $m\times n$ matrices with each entry in $\R_+$ by  $\R_+^{m\times n}$. We say such matrices are nonnegative and adopt the notation $A\geq {\bf 0}$.  If, in addition, $A $ has at least one positive entry, we say $A$ is positive ($A>{\bf 0}$). If all the entries are positive, we say $A$ is strictly positive ($A\gg {\bf 0}$). $A<{\bf 0}$ (or $\leq {\bf 0}$) if $-A>{\bf 0}$ ($\geq {\bf 0}$).  The nonnegativity, positivity, and strict positivity of vectors can be defined likewise.  For a square matrix $A$, $A \in \mathbb{H}$ means $A$ is Hurwitz, i.e., its  eigenvalues have negative real parts.
$A$ is  Metzler if the off-diagonal entries are nonnegative, which is equivalent to  $A \in \mathbb{M}$.  
$P\succ \, (\prec) \,  0$  means $P$ is a positive (negative) definite matrix, i.e. $x^\top P x > (<) \, 0$ for every $x\neq {\bf 0}$. 

Consider the following (time-varying) linear system:
\begin{align} \label{sys:def}
	\begin{split}
		y(t)&=C(t)x(t),\quad	\dot x(t)=A(t)x(t)+B(t)u(t) 
	\end{split}
\end{align}
where $x(t)$ is the $n$-dimensional state vector, $u(t)$ is the $p$-dimensional control input,  $y(t)$ is the $l$-dimensional output vector. Here, $A(t),\,B(t),\,C(t) $ are system matrices with compatible dimensions. We say this system is (internally) positive if for any nonnegative initial condition $x(0)\geq {\bf 0}$  and $u(t)\geq {\bf 0}$, it holds that $x(t)\geq {\bf 0}$ and $y(t)\geq {\bf 0}$ for $t\geq 0$.
It is said to be a Metzler system if $A(t)$ is Metzler and  $B(t),\, C(t)  \geq {\bf 0}$ for almost all $t \geq 0$.

Here is a lemma modified from \cite{angeli2003monotone}.
\begin{lemma}\label{lem:positive}
	If \eqref{sys:def} is a Metzler system then it is positive. Conversely, if \eqref{sys:def} is positive and $A(t)$, $B(t)$, $C(t)$  are continuous, then \eqref{sys:def} is a Metzler system.
\end{lemma}

\section{Problem Formulation}\label{sec:form}

In this paper, we consider a multi-agent system consisting of $N$ high-order dynamic agents of the following form:
\begin{align} \label{sys:follower}
	y_i(t)  = C_i x_i(t), \, {\dot x}_i(t) = A_i x_i(t) +B_i u_i(t) +D_id(t)
\end{align}
where $x_i(t) \in \mathbb{R}^{n_i}$,  $u_i(t) \in \mathbb{R}^{m_i}$, and $y_i(t) \in \mathbb{R}^{l}$ are the state, input, and output of agent $i=1,\, \dots,\, N $, while $d(t)\geq {\bf 0}\in \mathbb{R}^{q}$ represents some unmodeled time-varying external input (or disturbance) acting on agent $i$.   Here the external inputs are considered to be identical for each agent without loss of generality. Otherwise, we can lump all local external inputs together and redefine the associated input matrix $D_i$ for the same form \eqref{sys:follower}. Moreover, we assume matrix $A_i$ is Metzler, $B_i,\, C_i, \, D_i$ are nonnegative, and the disturbance $d(t)$ is locally essentially bounded.

We aim at effective controllers for the agents such that this multi-agent system can reach a positive output consensus specified by the following pattern:
\begin{align} \label{sys:leader}
	y_0(t)=C_0 x_0(t), \,  {\dot x}_0 (t)= A_0 x_0(t)
\end{align}
with internal state $x_0(t) \in \R^{n_0}$ and output $y_0(t) \in \R^{l}$. In other words, the output trajectory of each agent is expected to reach some solution to this differential equation simultaneously.  Since the agent is subject to external inputs, we expect that the patterned consensus error  converges to zero when $d(t)\equiv {\bf 0}$ while the influence of external inputs is attenuated to certain level when it is nonzero. 

Moreover, we are interested in distributed designs for this problem and assume the agents can share their own information with others. For this purpose, we utilize an undirected graph  $\G=\{\N,\,\mathcal{E},\,\mathcal{A} \}$ to represent the allowed information flow among them with node set $\N=\{1,\,\dots,\,N\}$, edge set $\mathcal{E}\subset \N\times \N$, and the symmetric adjacency matrix $\mathcal{A}=[a_{ij}]_{N\times N}$ (\cite{Mesbahi2010graph}). { When agents $i$ and $j$ can exchange information, there is an edge between them in this graph ${\G}$. For simplicity, we assume $a_{ij}=a_{ji}=1$ in this case and $a_{ij}=0$ otherwise. } Node $i$'s neighbor set is defined as $\mathcal{N}_i=\{j\mid (j,\, i)\in \mathcal{E} \}$. We denote $\mathcal{N}_i^0=\mathcal{N}_i\cup \{i\}$.  Moreover, we consider the case when the communication topology may be time-varying.  To describe the communication constraint precisely, we denote all possible communication graphs among these agents by $\{ \G _1,\,\dots,\, \G_p\}$ with $\mathcal{P}=\{1,\,2,\,\dots,\, p\}$. Consider a strictly increasing sequence of positive constants $\{t_\iota\}$ with $t_{0}=0$ and $\lim_{\iota\to \infty }t_{\iota}=\infty$. We suppose $t_{\iota+1}-t_{\iota}\geq \tau>0$ for any $\iota=0,\,1,\,\dots$ as that in \cite{hong2006tracking, ni2010leader}.  This sequence divides $[0,\,\infty)$ into some contiguous time intervals $[t_\iota,\, t_{\iota+1})$. Define a switching signal $\sigma(t)\colon [0, \, \infty) \rightarrow \mathcal{P}$. It is time-dependent and piece-wise constant. During each $[t_\iota,\, t_{\iota+1})$, all the agents can share their information according to graph $ \G_{\sigma({t})}$.

The considered robust positive consensus problem can be formulated as follows.
\begin{problem}\label{prob}
	Given a multi-agent system  \eqref{sys:follower}, graph $\G_p$, the consensus pattern \eqref{sys:leader}, and a constant $\gamma>0$, find distributed controllers of the following form:
	\begin{align}\label{ctrl:nominal}
		u_i&=f_i(t, x_j,\,\eta_j),\,\dot{\eta}_i=g_i(t,x_j,\,\eta_j),\, j \in \N_i^0(t)
	\end{align}
	with proper smooth functions $f_i$, $g_i$ and a  compensator $\eta_i\in \R^{n_{\eta_i}}$ such that, for any initial point $x_i(0)\geq {\bf 0}$, the closed-loop system \eqref{sys:follower} and \eqref{ctrl:nominal} satisfies the following properties.
	\begin{itemize}
		\item[1)] The trajectory of $x_i(t)$ is always nonnegative, i.e., $x_i(t)\geq {\bf 0} $ for any $t\geq 0$.
		\item[2)] It internally achieves a patterned consensus specified by system \eqref{sys:leader}. That is, there exists a positive constant $x_{00}\in \R^{n_0}_+$ such that,  $e_i(t) \triangleq y_i(t)-y_0(t)$ converges to ${\bf 0}$ as $t\to \infty$ with $y_0(t)=C_0 x_0(t)$ and $x_0(t)$ the corresponding trajectory of \eqref{sys:leader} starting from $x_0(0)=x_{00}$.
		\item[3)] The influence of external inputs is attenuated  such that the following inequality 		
		\begin{align*}
			\int_0^\infty \|e_i(s)\|^2 {\rm d}s \leq \gamma^2 \int_0^\infty \|d(s)\|^2{\rm d }s+ \kappa 
		\end{align*}
		holds for some positive constant $ \kappa$.
	\end{itemize}	
\end{problem}

The formulated problem has been partially discussed in the literature for standard nonpositive multi-agent systems under the name of output consensus or synchronization in \cite{ren2008consensus, xi2012output, Grip2012output}. Some recent attempts have been made in extending them to positive multi-agent systems assuming $A_0={\bf 0}$ \cite{valcher2013stabilizability, valcher2017consensus, liu2019positivity, bhattacharyya2022positive}. However, the obtained positive consensus results often require all agents share an identical high-order dynamics.  Here, we consider heterogeneous agent dynamics subject to external inputs and aim to ensure a nontrivial pattern consensus and  states' positivity simultaneously. 

Before the main results, we make  several extra assumptions to ensure the solvability of our problem as follows.

\begin{assumption} \label{ass:exo}
	Matrix $A_0$ is Metzler with   no  eigenvalues having  negative real parts.
\end{assumption}

\begin{assumption}\label{ass:graph}
	Each graph $\G_p$ is  connected.
\end{assumption}

\begin{assumption} \label{ass:regeq}
	For each $i=1, 2, \dots, N$, there exist constant matrices $X_{i}\in \R_+^{n_i\times n_0}$ and $U_i\in \R_+^{1\times n_i}$ such that
	\begin{align}\label{eq:regulator}
		\begin{split}
			A_i X_{i} +B_i U_i - X_{i} A_0 &={\bf 0}\\
			C_i X_{i} -C_0 &={\bf 0}
		\end{split}		
	\end{align}
\end{assumption}

Since the agents are positive, we assume the consensus pattern is also positive with nontrivial modes as stated in Assumption \ref{ass:exo}. {  Assumption \ref{ass:graph} is made to ensure the connectivity of the communication graphs. Under this assumption, the potential number of digraphs are finite and the Laplacian $L_p$ of each graph $\G_p$ is positive semidefinite with a simple zero eigenvalue. } Assumption \ref{ass:regeq} is known as the solvability of regulator equations for plant \eqref{sys:follower}  with an exosystem \eqref{sys:leader}  in the terminology of output regulation \cite{huang2004nonlinear}. Similar conditions have been widely used in the multi-agent literature, e.g., \cite{su2011cooperative, tang2016ijss, tang2018distributed}.

\section{Main Result} \label{sec:main} 

This section is devoted to the design of effective distributed controllers for each agent to solve our problem.  

\subsection{Two-step design scheme}

\begin{figure*}
	\centering 
	\includegraphics[width=0.55\textheight]{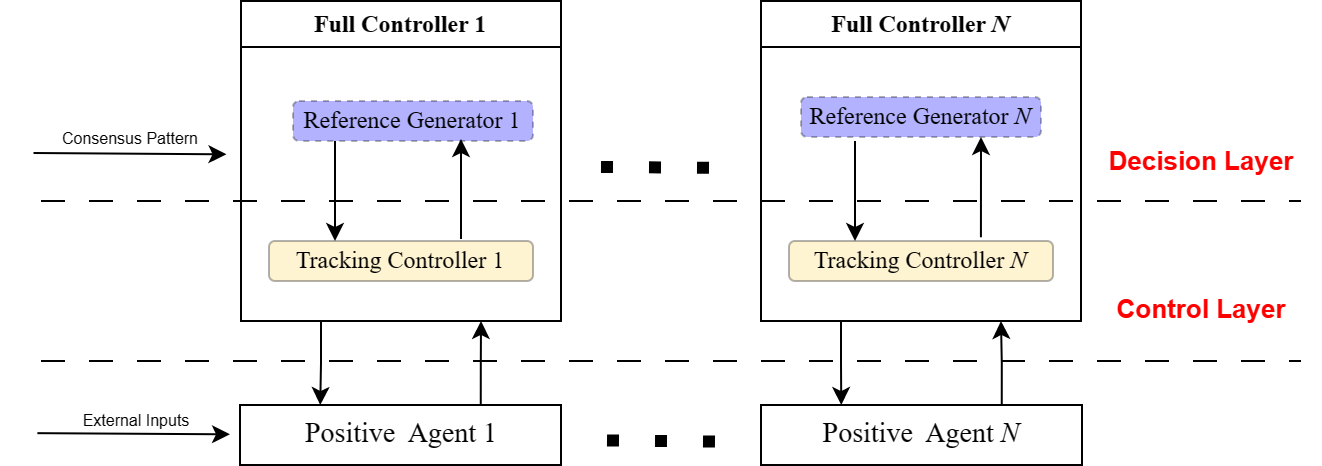}
	\caption{Illustration of two-step design scheme.}
	\label{fig:twosteps}
\end{figure*}

To hurdle the corresponding difficulties from agent dynamics and positive constraints, we present a two-step design to solve our problem. The basic layered structure is shown in Fig.~\ref{fig:twosteps}. We will first construct an auxiliary multi-agent system as local reference generators for each agent to meet the consensus pattern.  After that, we will focus on the resultant tracking problem for the original agents. In this way, the full controller for each agent consists of a reference generator and an effective tracking controller. 

According to our two-step scheme, we first present the following auxiliary multi-agent system with the same system matrix in pattern dynamics  \eqref{sys:leader}:
\begin{align}\label{sys:abs}
	\dot{w}_i=A_0w_i+{\I}_{n_0}v_i
\end{align}
where $w_i \in \R^{n_0}$ is the virtual state  and  $v_i$ the virtual input. Consider the typical distributed controller for agent
\eqref{sys:abs} as  $v_i=\mu \sum_{j=1}^N a_{ij}(t) (w_j-w_i)$.   The resultant reference generator is thus given as
\begin{equation} \label{sys:generator}
	\dot w_i = A_0 w_i + \mu \sum_{j=1}^N a_{ij}(t) (w_j-w_i),\quad i \in {\N}
\end{equation}
Let $w_{\rm av}(t)=\frac{\sum_{i=1}^N w_i(t)}{N}$, $\tilde w_i(t)=w_i(t)-w_{\rm av}(t)$, and $\tilde w =\mbox{col}(\tilde  w_1,\,\dots,\,\tilde  w_N)$. It can be verified that
\begin{align}\label{sys:average}
	\dot{w}_{\rm av}=A_0 w_{\rm av}
\end{align}
with an initial point $w_{\rm av}(0)=\frac{\sum_{i=1}^N w_i(0)}{N}$. That is, the trajectory $w_{\rm av}(t)$ is an admissible solution to the expected pattern dynamics \eqref{sys:leader}. 

Since there are a finite number of graphs fulfilling Assumption \ref{ass:graph},  $\underline{\lm} \triangleq \min\limits_{p\in \mathcal{P}} \{ \lm_2(L_p)\}$ is well-defined and strictly greater than $0$. Here is a key lemma on the performance of virtual positive multi-agent system  \eqref{sys:generator}.

\begin{lemma} \label{thm:1observer}
	Suppose Assumptions \ref{ass:exo}--\ref{ass:graph} holds. Let $\mu\geq \frac{\|A_0\|}{\underline{\lambda}}+1$. Then, along the trajectory of system \eqref{sys:generator},  it holds that $w_i(t)\geq  {\bf 0}$ and $\|\tilde{w}_i (t)\| \leq  \|\tilde w(0)\| e^{-\underline{\lambda} t}$ for any initial condition ${w}_i(0)\geq  {\bf 0}$.
\end{lemma}
\pb
	We first put \eqref{sys:abs} into a compact form:
	\begin{equation*}
		\dot{{w}} =(\I_N \otimes  A_0-\mu L_{\sigma(t)} \otimes \I_{n_0}) w %
	\end{equation*}
	where $w=\mbox{col}(  w_1,\,\dots,\,  w_N)$.  Since $A_0$ and $-L_p$ are Metzler for any $p\in \mathcal{P}$,  $(\I_N \otimes A_0-\mu L_p \otimes \I_{n_0})$ is also Metzler for any $\mu>0$. Then, we can conclude that the positivity of  $w(t)$ for $t\geq 0$ by Lemma \ref{lem:positive} from any $w_i(0)\geq {\bf 0}$.
	
	Next, we prove the positive consensus of $w_i(t)$. Recalling equation \eqref{sys:average}, it is sufficient to prove the exponential stability of the following error system at the origin:
	\begin{equation} \label{sys:error1w}
		\dot{\tilde{w}} =(\I_N \otimes  A_0 -\mu L_{\sigma(t)} \otimes \I_{n_0}) \tilde{w}
	\end{equation}
	For this purpose, we introduce two matrices $M_1\in \R^{N\times 1}$ and $M_2\in \R^{N\times (N-1)}$. Here, $M_1=\frac{\bf 1}{\sqrt{N}}$ and $M_2$ be a matrix such that $M_1^\top M_2={\bf 0}$, $M_2^\top M_2={\I}_{N-1}$, and $M_1 M_1^\top+M_2 M_2^\top={\I}_N$.
	Let $\check{w}_1=({M_1^\top }\otimes {\I}_{n_0} )\tilde w$ and $\check{w}_2=({M_2^\top }\otimes {\I}_{n_0})\tilde w$. It can be verified that $\dot{\check{w}}_1=(M_1^\top  \otimes  A_0 ) \tilde{w}={\bf 0}$ and $\check{w}_1(t)\equiv {\bf 0}$ by the definition of $w_{\rm av}(t)$ and the property  ${\bf 1}^\top L_p={\bf 0}$ under Assumption \ref{ass:graph}. This further implies that
	\begin{align*} 
		\dot{\check{w}}_2
		&=[\mathbb{I}_{N-1}\otimes A_0-\mu (M_2^\top L_{\sigma(t)} M_2) \otimes \I_{n_0}]\check w_2
	\end{align*}
	It can be verified that the matrix $M_2^\top L_{\sigma(t)} M_2$ is positive definite with only real eigenvalues $0<\lm_2(L_{\sigma(t)})\leq \dots\leq \lm_N(L_{\sigma(t)})$ under Assumption \ref{ass:graph}.
	
	Let us choose a candidate of common Lyapunov function $V_{\tilde{w}}=\frac{1}{2}\tilde{w}^\top \tilde{w}$ for the switched positive linear system \eqref{sys:error1w}.  
	It can be found that $V_{\tilde {w}}=\frac{1}{2}\check{w}_1^\top \check{w}_1+\frac{1}{2}\check{w}_2^\top \check{w}_2=\frac{1}{2}\check{w}_2^\top \check{w}_2$ along the  trajectory of  system \eqref{sys:error1w}. We assume $\sigma(t)=p$ during $[t_k,\,t_{k+1})$.
	Since $M_2^\top L_{p} M_2$ is positive definite, there must be a unitary matrix $U_p\in \R^{(N-1)\times (N-1)}$ such that $U_p^{\top} [M_2^\top L_{\sigma(t)} M_2] U_p =D_p$ where $D_p$ is a diagonal matrix with the eigenvalues $\lm_2(L_{p}),\, \dots,\, \lm_N(L_{p})$ on the diagonal. Let $\bar {\check w}_2=(U_p^\top \otimes {\mathbb I}_{n_0})\check w_2$. During this time interval, we have
	\begin{align*}
		\dot{\bar {\check w}}_2=(\I_{N-1} \otimes  A_0-\mu D_P \otimes \I_{n_0}) {\bar {\check w }}_2
	\end{align*}
	Note that $\I_{N-1} \otimes  A_0 -\mu D_P \otimes \I_{n_0}$ is a block diagonal matrix with diagonal blocks of the form $A_0-\mu \lambda_i(L_p){\mathbb I}_{n_0}$ for $i=2,\,\dots,\,N$. Jointly using the fact that $\check{w}_2^\top \check{w}_2={\bar {\check{w}}}_2^\top \bar {\check{w}}_2$, we set $\mu$ as in the theorem condition and obtain 
	\begin{align*}
		\dot{V}_{\tilde w}&={\bar {\check w}}_2^\top (\I_{N-1} \otimes  A_0 -\mu D_P \otimes \I_{n_0}) {\bar {\check w}}_2\\
		&\leq (\|A_0\|-\mu {\lambda}_2(L_p))\|\check w_2\|^2 \leq -2 \underline{\lambda} V_{\tilde{w}}
	\end{align*}
	It is verified that this inequality holds for any $p\in \mathcal{P}$ and thus holds over $[0,\,\infty)$.
	According to Theorem 3.1 in \cite{blanchini2015switched},
	$\tilde w_i(t)$ exponentially converges to zero as $t\to \infty$. Solving this inequality gives ${V}_{\tilde w}(t)\leq  {V}_{\tilde w}(0)e^{-2\underline{\lambda} t}$. It follows then
	\begin{align*}
		\|\tilde{w}_i(t)\|\leq \|\tilde{w} (t)\|=\sqrt{2 {V}_{\tilde w}(t)}\leq  \|\tilde w(0)\| e^{-\underline{\lambda} t}
	\end{align*} 
	The proof is thus completed. 
\pe

Lemma \ref{thm:1observer} is motivated by the observer-based designs in \cite{su2011cooperative,tang2016ijss}. With this lemma, we will embed the auxiliary generator into some reference tracking controller for agent \eqref{sys:follower} to solve the formulated positive patterned consensus problem in the next subsection.

\subsection{Algorithm design and solvability analysis}

In this subsection, we combine the reference generator \eqref{sys:generator} with a robust tracking controller towards the final distributed controller for multi-agent system \eqref{sys:follower}.

Since each agent has positive linear dynamics, we consider distributed controllers of the following form:
\begin{align}\label{con-obs-1}
	\begin{split}
		u_i&=K_{1i} x_i + K_{2i} w_i\\
		\dot{w}_i &= A_0 w_i + \mu \sum_{j=1}^N a_{ij}(t) (w_j-w_i),\quad i \in  {\N}
	\end{split}
\end{align}
where  $\mu$ is chosen as above and $K_{1i}$, $K_{i2}$ are matrices to be specified later. 

Here is the main result of this paper.

\begin{theorem} \label{th:L2state}
	Suppose Assumptions \ref{ass:exo}--\ref{ass:regeq} hold. The formulated robust positive consensus problem with a given pattern \eqref{sys:leader} and performance level $\gamma>0$ is solved by a distributed controller of the form \eqref{con-obs-1} if there exist diagonal matrices $Q_1\,\dots,\,Q_N>{\bf 0}$ and a scalar $\delta>0$ such that the following inequalities hold:
	\begin{align}  \label{eq:robust1}
		&A_iQ_i-B_{i}B_i^\top+  \delta Q_i >{\bf 0}\\ 
		&Q_i A_i^\top + A_iQ_i-2B_iB_i^\top+ \frac{1}{\gamma^2}D_iD_i^\top + Q_i C_i^\top C_iQ_i \prec {\bf 0} \nonumber
	\end{align}
\end{theorem}

\pb  We attach the controller \eqref{con-obs-1} to agent \eqref{sys:follower} and obtain the following composite system:
\begin{align} \label{sysclosedloop}
	\begin{split}
		\dot{x}_i &= (A_i+B_iK_{1i})x_i+ B_iK_{2i} w_i+D_i d(t)\\
		\dot{w}_i &= A_0 w_i + \mu \sum_{j=1}^N a_{ij}(t) (w_j-w_i)\\
		y_i&= C_i x_i,\quad i \in  {\N}
	\end{split}
\end{align}	
To ensure the solvability of our problem, we choose $K_{1i}=-B_i^\top Q_i^{-1}$ and $K_{2i}=U_i-K_{1i}X_i$ with $Q_i$, $U_i$, and $X_i$ defined as above. We are going to verify the three required properties in the formulation section.   

We first show the closed-loop system is indeed positive. In fact, according to \eqref{eq:robust1}, the matrix $\bar A_i=A_i+B_iK_{1i}$ is Metzler by definition.  Meanwhile, the matrices $-K_{1i}$ and $K_{2i}$ can be found to be nonnegative. Hence, by Lemma \ref{lem:positive}, the ${x}_i$-subsystem is positive with $w_i$ and $d(t)$ as its input. Since $w_i(t)>{\bf 0}$ by Lemma \ref{thm:1observer}, we have that from any $x_i(0)>{\bf 0}$,  $x_i(t)>{\bf 0}$ holds for any $t\geq 0$.  

Next, we verify the second property in our formulation. { We suppose $d(t)\equiv {\bf 0}$ and show that there exists some $x_{00} \geq {\bf 0}$ such that $e_i(t)$ converges to zero as $t\to \infty$ in this case.} That is, the patterned consensus is indeed reached under this controller \eqref{con-obs-1}. To this end, we can verify that
\begin{equation*}
	\begin{split}
		Q_i	\bar A_i^\top + \bar A_iQ_i&=Q_i(A_i-B_iB_i^\top Q_i^{-1})^\top+ (A_i-B_iB_i^\top Q_i^{-1})Q_i\\
		&=Q_iA^\top +A_i Q_i-2B_iB_i^\top \prec {\bf 0}
	\end{split}
\end{equation*}
Thus $\bar A_i$ is Hurwitz. Moreover, one can further determine some $c_0>0$ such that $Q_i	\bar A_i^\top + \bar A_iQ_i\prec -c_0Q_i^2$ holds for any $i$.  By our two-step design procedure and Lemma \ref{thm:1observer}, we set $x_{00}=w_{\rm av}(0)$. Then, $e_i(t)=C_i x_i(t)-C_0 w_{\rm av}(t)$.  Performing the coordinate transformation $\tilde x_i=x_i-X_iw_{\rm av}$, we have
\begin{align}\label{sys:error-final-1}
	\begin{split}
		\dot{\tilde x}_i &= (A_i+B_iK_{1i})\tilde x_i+ B_iK_{2i} \tilde w_i\\
		\dot{\tilde w} &=(\I_N \otimes  A_0 -\mu L_{\sigma(t)} \otimes \I_{n_0}) \tilde{w}\\
		e_i&=C_i\tilde x_i
	\end{split}
\end{align}

Choose $V_i(t)={\tilde x}_i^\top(t) Q_i^{-1} {\tilde x}_i(t)+ \iota_i V_{\tilde w}(t)$ with a constant $\iota_i>0$ to be determined later. It is positive definite in $[\tilde x_i^\top(t),\,\tilde w(t)^\top]^\top$. 
Recalling the proofs in Lemma \ref{thm:1observer}, we take its time derivative along the trajectory of \eqref{sys:error-final-1} and obtain
\begin{align*}
	\dot{V}_i(t)&=2{\tilde x}_i^\top Q_i^{-1} [(A_i+B_iK_{1i})\tilde x_i+  B_iK_{2i} \tilde w_i]+ \iota_i\dot{V}_{\tilde w}\\
	&\leq 2{\tilde x}_i^\top Q_i^{-1}(A_iQ_i-B_i B_i^\top) Q_i^{-1}\tilde x_i\\
	& + 2\tilde x_i^\top Q_i^{-1} B_iK_{2i} \tilde w_i-2\iota_i \underline{\lambda} V_{\tilde w}\\
	&\leq -c_0 \|{\tilde x}_i\|^2+ 2\|Q_i^{-1} B_iK_{2i} \|  \|\tilde x_i\| \|\tilde w_i\|-2\iota_i \underline{\lambda} V_{\tilde w}
\end{align*}
We complete the square and have
\begin{align*}
	\dot{V}_i(t)
	&\leq -\frac{c_0}{2}\|{\tilde x}_i\|^2-( \iota_i\underline{\lm}-\frac{2}{c_0}\|Q_i^{-1} B_iK_{2i}\|^2)\|\tilde w\|^2 
\end{align*}
Letting $\iota_i\geq \frac{2}{\underline{\lm}}\max\{ \frac{2}{c_0}\|Q_i^{-1}B_iK_{2i}\|^2,\,1\}$ implies
\begin{align*}
	\dot{V}_i(t)&\leq -\frac{c_0}{2}\|{\tilde x}_i\|^2-\|{\tilde w}\|^2
\end{align*}
According to Theorem 3.1 in \cite{blanchini2015switched},   $V_i(t)$ and $\tilde x_i (t)$ both exponentially converge to $0$ as $t\to \infty$.
As a result, the tracking error $e_i(t)=C_i\tilde x_i(t)$ also converges to ${\bf 0}$ exponentially fast. This means that the agent outputs indeed reach a positve consensus and the consensus confirms the pattern specified by \eqref{sys:leader} when the external input $d(t)$ vanishes.

To confirm the  third property, we use the same Lyapunov function $V_i$ and take its time derivative along the trajectory of system \eqref{sys:error-final-1}. It follows then 
\begin{equation*}
	\begin{split}
		\dot  V_i&=2{\tilde x}_i^\top Q_i^{-1} [(A_i+B_iK_{1i})\tilde x_i+  B_iK_{2i} \tilde w_i+E_id(t)]+ \iota_i\dot{V}_{\tilde w}\\
		&\leq 2{\tilde x}_i^\top Q_i^{-1}(A_iQ_i-B_i B_i^\top) Q_i^{-1}\tilde x_i\\ 
		&+ 2\tilde x_i^\top Q_i^{-1} B_iK_{2i} \tilde w_i+ 2\tilde x_i^\top Q_i^{-1} E_id(t)-2\iota_i \underline{\lambda} V_{\tilde w}
	\end{split}
\end{equation*}
Under the theorem conditions, there exists some small constant  $\tilde c_0>0$ such that the following inequality holds:
\begin{equation*}
	Q_i A_i^\top + A_i Q_i- 2B_iB_i^\top+\frac{1}{\gamma^2}E_iE_i^\top +  Q_i C_i^\top C_iQ_i \prec -\tilde c_0 Q_i^2
\end{equation*}
With this inequality, we further have
\begin{align*}
	\dot  V_i(t)
	&\leq -\tilde c_0 \|{\tilde x}_i\|^2-\tilde x_i^\top Q_i^{-1}(\frac{1}{\gamma^2}E_iE_i^\top + Q_i C_i^\top C_iQ_i ) Q_i^{-1}\tilde x_i\\ 
	&+2\tilde x_i^\top Q_i^{-1} B_iK_{2i} \tilde w_i
	+ 2\tilde x_i^\top Q_i^{-1} E_id(t)-2\iota_i \underline{\lambda} V_{\tilde w}\\
	&\leq -\frac{\tilde c_0}{2} \|{\tilde x}_i\|^2-(\iota_i \underline{\lm}- \frac{2}{\tilde c_0}\|Q_i^{-1} B_iK_{2i}\|^2) \|\tilde w\|^2\\ 
	&+\gamma^2 \|d(t)\|-\|e_i\|^2
\end{align*}
By letting $\iota_i\geq \frac{2}{\underline{\lm}} \max\{\frac{2}{\tilde c_0} \|Q_i^{-1} B_iK_{2i}\|^2,\,1\}$, we have
\begin{align*}
	\dot  V_i(t)&\leq -\frac{\tilde c_0}{2} \|{\tilde x}_i\|^2- \|\tilde w\|^2+\gamma^2 \|d(t)\|-\|e_i\|^2
\end{align*}
Integrating this inequality from $0$ to $\infty$ yields that
\begin{align*}  
	V_i(\infty)-V_i(0)\leq \int_{0}^{\infty}  \gamma^2\|d(s)\|^2{\rm d}s -\int_{0}^{\infty} \|e_i(s)\|^2 {\rm d}s
\end{align*}
Using the fact that $V_i(t)\geq 0$ for any $t$, we have 
\begin{align*}  
	\int_{0}^{\infty} \|e_i(s)\|^2 {\rm d}s\leq 	\int_{0}^{\infty}  \gamma^2\|d(s)\|^2{\rm d}s+V_i(0) 
\end{align*}
The proof is thus complete.  \pe

When  $d(t)\equiv {\bf 0}$, condition \eqref{eq:robust1} can be further relaxed as follows to ensure a patterned positive consensus:
\begin{align}  \label{eq:LP}
	\begin{split}
		A_iQ_i-B_{i}B_i^\top  +  \delta Q_i >0\\
		Q_iA^\top +A_i Q_i-2B_iB_i^\top \prec 0
	\end{split}
\end{align}

We summarize the result as follows.
\begin{corollary} \label{th:state}
	Suppose Assumptions \ref{ass:exo}--\ref{ass:regeq} hold and there exist diagonal matrices $Q_1\,\dots,\,Q_N>{\bf 0}$ and a scalar $ \delta >0$ fulfilling \eqref{eq:LP}. Then, under the following controller
	\begin{align}\label{con-obs-11}
		\begin{split}
			u_i&=-B_i^\top Q_i^{-1}x_i + [U_i+B_i^\top Q_i^{-1}X_i] w_i\\
			\dot{w}_i &= A_0 w_i + \mu \sum_{j=1}^N a_{ij}(t) (w_j-w_i),\quad i \in  {\N}
		\end{split}
	\end{align}
	the trajectory of $x_i(t)$ is always nonnegative and the multi-agent system \eqref{sys:follower} internally achieves a patterned consensus specified by  \eqref{sys:leader}. 
\end{corollary}

In many circumstances, the state $x_i$ may not be available for us. Here we present an output feedback extension for \eqref{con-obs-1} as follows to solve our problem:
\begin{align} \label{con-obs-2}
	\begin{split}
		u_i&=K_{1i} \xi_i  + K_{2i} w_i  \\
		\dot{\xi}_i &=(A_i- K_{3i}C_i) \xi_i+B_iu_i +K_{3i} y_i \\
		\dot{w}_i &= A_0  w_i  +  \mu \sum_{j=1}^N a_{ij}(t) (w_j-w_i),\quad i \in {\N}
	\end{split}
\end{align}
where $K_{3i} \in \mathbb{R}^{n_i \times l}$ is a chosen gain matrix to be specified later. To meet the positivity requirement, we set $w_i(0)={\bf 0}$ for $i=1,\,\dots,\,N$ as in \cite{roszak2009necessary}.  


\begin{theorem} \label{thm:output}  
	Suppose Assumptions \ref{ass:exo}--\ref{ass:regeq} hold. The formulated robust positive consensus problem with a given pattern \eqref{sys:leader} and performance level $\gamma>0$ is solved by a distributed controller of the form \eqref{con-obs-2} if there exist diagonal matrices  $P_i,\, Q_i>{\bf 0}$ and a  scalar   $ \delta>0$  such that the following inequalities hold:
	\begin{align*} 
		&P_i A_i-C_{i}^\top C_i  + \delta P_i >{\bf 0}\\		
		&A_i^\top P_i+P_iA_i-2C_i^\top C_i \prec {\bf 0}\\ 				
		&A_iQ_i-B_{i}B_i^\top+ \delta Q_i >{\bf 0}\\
		&Q_i A_i^\top + A_iQ_i-2B_iB_i^\top+ \frac{1}{\gamma^2}D_iD_i^\top + Q_i C_i^\top C_iQ_i \prec {\bf 0}
\end{align*}
\end{theorem}
\pb The proof is analogous to that of Theorem \ref{th:L2state} and we only give a sketch to save space. 
Let $K_{3i}=P_i^{-1}C_i^\top$. It is verified that $A_i-K_{3i}C_i$ is Hurwitz and Metzler. It follows 
\begin{align} \label{sysclosed2}
	\begin{split}
		\dot{x}_i &= (A_i+B_iK_{1i})x_i-B_iK_{1i}\bar x_i+B_iK_{2i} w_i+E_id(t)\\
		\dot{\bar x}_i&=(A_i-K_{3i}C_i)\bar x_i\\
		y_i&=C_ix_i ,\quad i \in  {\N}
	\end{split}
\end{align}
where  $\bar x_i=x_i-\xi_i$. Since $\bar x_i(0)\geq {\bf 0}$, $\bar x_i(t)\geq {\bf 0}$ for any $t$. This further implies the nonnegativity of $-B_iK_{1i}\bar x_i+B_iK_{2i} w_i+E_id(t)$. Consequently, we have $x_i(t)\geq {\bf 0}$ for any $t\geq 0$. 

To show the rest two properties, we focus on the following error system: 
\begin{align}\label{sys:error-final-3}
	\begin{split}
		\dot{\tilde x}_i  &= (A_i+B_iK_{1i})\tilde x_i -B_iK_{1i}\bar x_i+ B_iK_{2i} \tilde w_i  +E_id(t)\\
		\dot{\bar x}_i &=(A_i-K_{3i}C_i)\bar x_i \\
		\dot{\tilde w}  &=(\I_N \otimes  A_0 -\mu L_{\sigma(t)} \otimes \I_{n_0}) \tilde{w} \\
		e_i &=C_i\tilde x_i 
	\end{split}
\end{align}
Following similar arguments in the proof of Theorem \ref{th:L2state}, one can further determine constant $c_1>0$ such that 
$\hat{A}_i^\top P_i +P_i	\hat{A}_i \prec -c_1 \mathbb{I}_{n_i}$ holds for any $  i \in  {\N}$.  
Then  we take $V_i(t)={\tilde x}_i^\top(t) Q_i^{-1} {\tilde x}_i(t)+ \iota_i V_{\tilde w}(t)+m_i\bar{x}_i^\top(t) P_i \bar{x}_i(t)$ with $\iota_i$ and $m_i$ to be specified later. Using the theorem conditions, we can obtain that its derivative along system \eqref{sys:error-final-3} satisfies
\begin{align*}
	\dot  V_i(t)
	&\leq -\tilde c_0 \|{\tilde x}_i\|^2 -\tilde x_i^\top Q_i^{-1}(\frac{1}{\gamma^2}E_iE_i^\top + Q_i C_i^\top C_iQ_i ) Q_i^{-1}\tilde x_i\\ 
	&+2\tilde x_i^\top Q_i^{-1} B_iK_{2i} \tilde w_i -2\tilde x_i^\top Q_i^{-1} B_iK_{1i} \bar x_i\\
	&+ 2\tilde x_i^\top Q_i^{-1} E_id(t)-2\iota_i \underline{\lambda} V_{\tilde w}-m_ic_1 \|\bar{x}_i\|^2
\end{align*}
Completing the squares implies 
\begin{align*}
	\dot  V_i(t)&\leq -\frac{\tilde c_0}{2} \|{\tilde x}_i\|^2-(\iota_i \underline{\lm}- \frac{4}{\tilde c_0}\|Q_i^{-1} B_iK_{2i}\|^2) \|\tilde w\|^2\\
	&-(m_ic_1-\frac{4}{\tilde c_0}\|Q_i^{-1} B_iK_{1i}\|^2 ) \|\bar{x}_i\|^2\\
	& +\gamma^2 \|d(t)\|-\|e_i\|^2
\end{align*}
By letting $\iota_i\geq \frac{2}{\underline{\lm}} \max\{\frac{4}{\tilde c_0} \|Q_i^{-1} B_iK_{2i}\|^2,\,1\}$ and 
$m_i\geq \frac{2}{c_1}\max\{\frac{4}{\tilde c_0} \|Q_i^{-1} B_iK_{1i}\|^2,\,1\}$, we have
\begin{align*}
	\dot  V_i(t)&\leq -\frac{\tilde c_0}{2} \|{\tilde x}_i\|^2- \|\bar{x}_i\|^2- \|\tilde w\|^2+\gamma^2 \|d(t)\|-\|e_i\|^2
\end{align*}
Integrating this inequality from $0$ to $\infty$ yields that
\begin{align*}  
	V_i(\infty)-V_i(0)\leq \int_{0}^{\infty}  \gamma^2\|d(s)\|^2{\rm d}s -\int_{0}^{\infty} \|e_i(s)\|^2 {\rm d}s
\end{align*}
Using the fact that $V_i(t)\geq 0$ for any $t$, we have 
\begin{align*}  
	\int_{0}^{\infty} \|e_i(s)\|^2 {\rm d}s\leq 	\int_{0}^{\infty}  \gamma^2\|d(s)\|^2{\rm d}s+V_i(0) 
\end{align*}
The proof is thus complete. 
\pe

\begin{remark} 
	Compared with most existing positive consensus results \cite{valcher2017consensus, wu2018observer, liu2019positivity, yang2019positive, bhattacharyya2022positive}, we present a two-step design scheme to solve the problem. Although the controller \eqref{con-obs-2}, particularly the local reference part,  reduces to the observer-based type of control laws as that in \cite{yang2019positive, bhattacharyya2022positive}, we are able to handle heterogeneous positive multi-agent systems whose dynamics can be different from each other in both system matrices and orders over switching communication topologies.
	Moreover, the expected consensus trajectory of the multi-agent system is allowed to be of a more general prespecified pattern including nonnegative static consensus in \cite{bhattacharyya2022positive} as a special case.
\end{remark} 

\begin{remark}
	It is interesting to remark that the presented algorithms are mainly built upon several matrix inequalities, which can be taken as  positive counterparts of $\gamma$-suboptimal $\mathcal{H}_\infty$ design for similar problems \cite{li2015stability,xu2022positive}.  In practice, we can convert them to linear ones and then solve them using standard numerical softwares.
\end{remark}

\section{Simulation}  \label{sec:simulation}

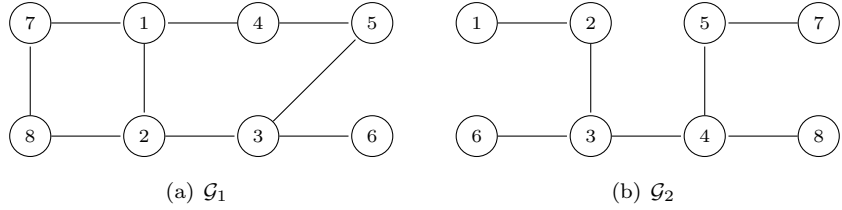
\begin{figure*}
\centering
\subfigure[$\mathcal{G}_1$]
{
	\begin{tikzpicture}[shorten >=1pt, node distance=1.5 cm, >=stealth',
		every state/.style ={circle, minimum width=0.3cm, minimum height=0.3cm}]
		\node[align=center,state](node1) {\scriptsize 1};
		\node[align=center,state](node2)[below of=node1]{\scriptsize 2};
		\node[align=center,state](node3)[right of=node2]{\scriptsize 3};
		\node[align=center,state](node4)[above of=node3]{\scriptsize 4};
		\node[align=center,state](node5)[right of=node4]{\scriptsize 5};
		\node[align=center,state](node6)[right of=node3]{\scriptsize 6};
		\node[align=center,state](node7)[left of=node1]{\scriptsize 7};
		\node[align=center,state](node8)[left of=node2]{\scriptsize 8};
		\path[-] (node1) edge (node2)
		(node2) edge (node3)
		(node4) edge (node1)
		(node4) edge (node5)
		(node3) edge (node6)
		(node3) edge (node5)
		(node7) edge (node1)
		(node8) edge (node2)
		(node8) edge (node7);
\end{tikzpicture}}\qquad
\subfigure[$\mathcal{G}_2$]
{
	\begin{tikzpicture}[shorten >=1pt, node distance=1.5 cm, >=stealth',
		every state/.style ={circle, minimum width=0.3cm, minimum height=0.3cm}]
		\node[align=center,state](node1) {\scriptsize 1};
		\node[align=center,state](node2)[right of=node1]{\scriptsize 2};
		\node[align=center,state](node3)[below of=node2]{\scriptsize 3};
		\node[align=center,state](node4)[right of=node3]{\scriptsize 4};
		\node[align=center,state](node5)[above of=node4]{\scriptsize 5};
		\node[align=center,state](node6)[below of=node1]{\scriptsize 6};
		\node[align=center,state](node7)[right of=node5]{\scriptsize 7};
		\node[align=center,state](node8)[right of=node4]{\scriptsize 8};
		\path[-] (node1) edge (node2)
		(node2) edge (node3)
		(node3) edge(node4)
		(node4) edge (node5)
		(node6) edge (node3)
		(node7) edge (node5)
		(node8) edge (node4);
\end{tikzpicture}}
\caption{The communication graphs in our example.}\label{fig:graph}
\end{figure*}

In this section, we consider an eight-agent system to illustrate the effectiveness of our preceding designs. 

Suppose the system matrices are as follows:
\begin{align*}
A_i&=\begin{bmatrix} -2&1&1\\1&-3&0\\1&1&-1
\end{bmatrix},\, B_i=\begin{bmatrix}
	0\\0\\1
\end{bmatrix},\, C_i=\begin{bmatrix}
	0\\0\\1
\end{bmatrix}^\top,\, D_i=\begin{bmatrix}
	1\\0\\ 0
\end{bmatrix}, \, i=1,\,2,\,7
\end{align*}
and
\begin{align*}
A_i&=\begin{bmatrix}-2&1\\ 0&0
\end{bmatrix},\, B_i=\begin{bmatrix}
	0\\1
\end{bmatrix},\, C_i=\begin{bmatrix}
	0\\1
\end{bmatrix}^\top,\, D_i=\begin{bmatrix}
	1\\1
\end{bmatrix}, \, i=3,\,4
\end{align*}
and 
\begin{align*}
A_i&=\begin{bmatrix} 0&0 \\ 1&-3
\end{bmatrix},\, B_i=\begin{bmatrix}
	1\\0
\end{bmatrix},\, C_i=\begin{bmatrix}
	2\\ 0
\end{bmatrix}^\top, \, D_i=\begin{bmatrix}
	1\\ 1
\end{bmatrix}, \, i=5,\,6,\,8
\end{align*}
It can be verified by Lemma \ref{lem:positive} that these agents are indeed positive. Suppose the  communication graph is alternatively switching between $\G_1$ and $\G_2$ given in Fig.~\ref{fig:graph}  every $10$ seconds.  Assumption \ref{ass:graph} is then fulfilled. 

Consider a consensus pattern with
\begin{align} \label{sys:simu}
A_0=\begin{bmatrix}
	0.01&0.01\\0&0
\end{bmatrix},\quad  C_0=\begin{bmatrix}
	1\\ 1
\end{bmatrix}^\top
\end{align}
Solving the regulator equations \eqref{eq:regulator} gives
\begin{align*}
&	X_i=\begin{bmatrix}
	0.5960 & 0.5960\\
	0.1980  & 0.1980\\
	1 & 1 
\end{bmatrix},  U_i=\begin{bmatrix}
	0.2160 \\ 0.2160
\end{bmatrix}^\top, \, i=1,\,2,\,7\\
&	X_i=\begin{bmatrix}
	0.4975 &  0.4975\\
	1.0000 & 1.0000
\end{bmatrix},  U_i=\begin{bmatrix}
	0.0100 \\ 0.0100
\end{bmatrix}^\top,\, i=3,\,4\\
&	X_i=\begin{bmatrix}
	0.5000 &  0.5000\\
	0.1661 & 0.1661
\end{bmatrix},  U_i=\begin{bmatrix}
	0.0050 \\ 0.0050
\end{bmatrix}^\top,\, i=5,\,6,\,8	
\end{align*}
Thus Assumption \ref{ass:regeq} is confirmed. We employ the output feedback controller \eqref{con-obs-2} to solve our problem with $\gamma=4$.

\begin{figure*}
	\centering
	\subfigure[$x_1$, $x_3$ and $x_5$]
	{\includegraphics[width=0.7\linewidth]{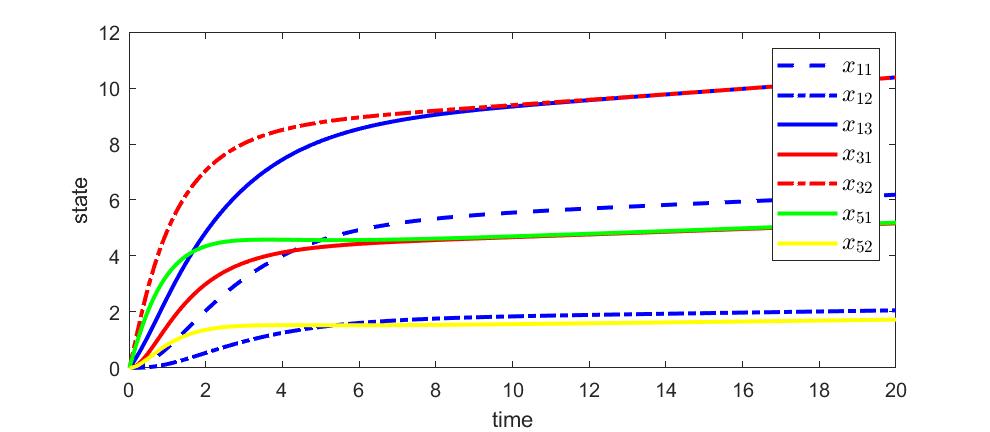}}\\
	\subfigure[$y_i$]	
	{\includegraphics[width=0.7\linewidth]{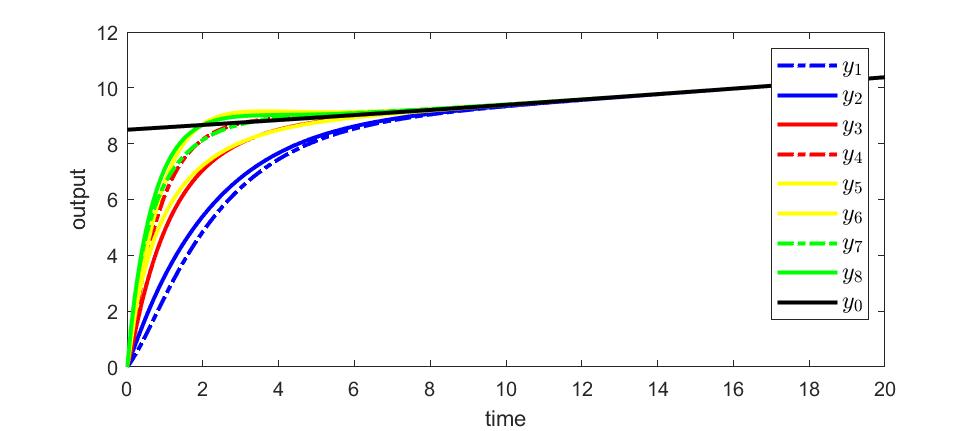}}
	\caption{Performance of \eqref{con-obs-2} for positive consensus.}
	\label{fig:simu:state:output}
\end{figure*}

For this purpose, we solve the inequalities in Theorem \ref{thm:output} and obtain the following gain matrices:
$K_{11}=K_{12}=K_{17}=[0~0~-1]$, $K_{13}=K_{14}=[0~-1]$, $K_{15}=K_{16}=K_{18}=[-1~0]$,
$K_{21}=K_{22}=K_{27}=[1.2160~1.2160]$, $K_{23}=K_{24}=[1.0100~1.0100]$, $K_{25}=K_{26}=K_{28}=[0.5050~ 0.5050]$,
$K_{31}=K_{32}=K_{37}=[0~0~1]^\top$, $K_{33}=K_{34}=[0~1]^\top$ and $K_{35}=K_{36}=K_{38}=[1~ 0]^\top$. Let $w_i(0)=\mbox{col}(i-0.5,\,i)$, $\xi_i(0)={\bf 0}$. Choose $\mu=3$ for the generator and all other initials are randomly generated between $0$ and $7$,

\begin{figure*}
	\centering
	\subfigure[$x_1$, $x_3$ and $x_5$]
	{\includegraphics[width=0.7\linewidth]{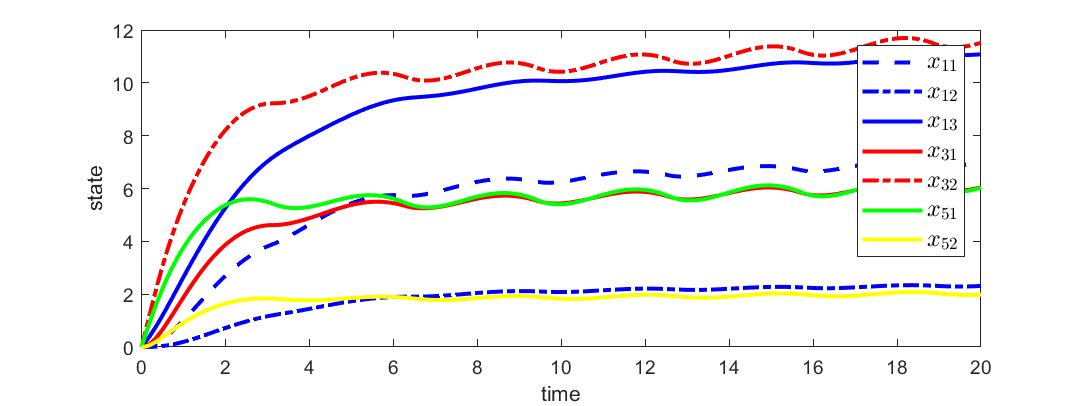}}\\
	\subfigure[$L_2$ gain robust performance of \eqref{con-obs-2}]	
	{\includegraphics[width=0.7\linewidth]{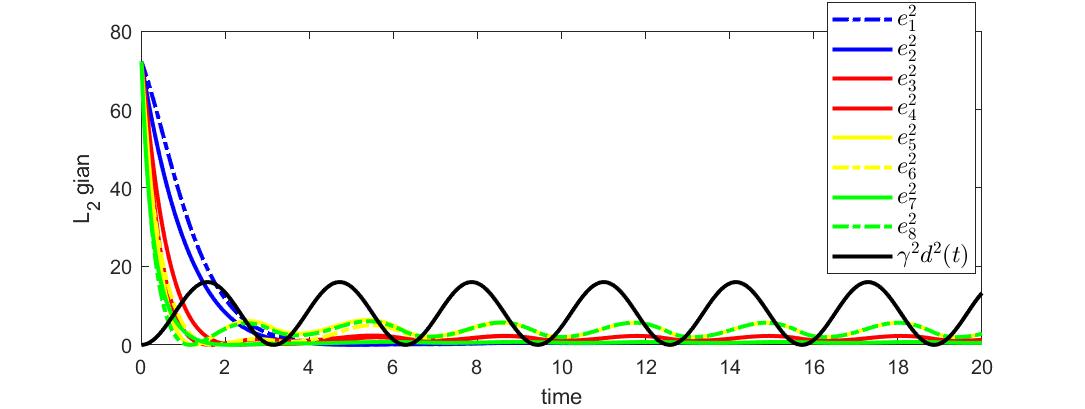}}
	\caption{State and $L_2$ gain robust performance of \eqref{con-obs-2}.} \label{fig:simu:L_2:output}
\end{figure*}

We first set $d(t)\equiv 0$ and obtain an exact positive consensus as depicted in Fig.~\ref{fig:simu:state:output}. Then, let $d(t)={|\sin(0.01t)|}$,  the state and $L_2$ gain performance  of  controller \eqref{con-obs-2} can be found in Fig.~\ref{fig:simu:L_2:output}.
It can be found that all agents reach a robust positive consensus even the expected consensus trajectory tends to be unbounded in this case. At the same time,   the components of $x_1(t)$, $x_3(t)$ and $x_5(t)$ are observed to stay in the positive orthant. These observations verify the effectiveness of  \eqref{con-obs-2} to solve the robust positive   consensus problem for heterogeneous multi-agent system \eqref{sys:follower}.

\section{Conclusion}\label{sec:conclusion}

We have formulated and solved the robust positive consensus problem for a group of high-order positive multi-agent systems with external inputs. To handle the positive constraint and heterogeneous and uncertain agent dynamics, we have proposed a two-step design method and finally developed two different kinds of  effective rules for these agents to attain a robust consensus having the expected dynamic pattern while their states fulfill the positive constraints  even under switching communication topologies. { In the future, we may consider the same problem but for uncertain nonlinear positive multi-agent systems with more general communication graphs.}

%
%
%


\end{document}